\documentclass[conference]{IEEEtran}
\IEEEoverridecommandlockouts
\usepackage{cite}
\usepackage{amsmath,amssymb,amsfonts}
\usepackage{algorithmic}
\usepackage{graphicx}
\usepackage{textcomp}
\usepackage{xcolor}
\def\BibTeX{{\rm B\kern-.05em{\sc i\kern-.025em b}\kern-.08em
    T\kern-.1667em\lower.7ex\hbox{E}\kern-.125emX}}

\makeatletter
\newcommand{\linebreakand}{%
  \end{@IEEEauthorhalign}
  \hfill\mbox{}\par
  \mbox{}\hfill\begin{@IEEEauthorhalign}
}
\makeatother
\author{
\IEEEauthorblockN{Fang Liu}
\IEEEauthorblockA{
    \textit{Yale University} \\
    Forest Hills,NY,11375, USA\\
    fangliu435@gmail.com}
\and
\IEEEauthorblockN{Shaobo Guo}
\IEEEauthorblockA{
    \textit{Nankai University} \\
    Chicago, IL 60616, USA \\
    seanguo2017@gmail.com}
\and
\IEEEauthorblockN{Qianwen Xing}
\IEEEauthorblockA{
    \textit{ The University of Chicag} \\
    Chicago, IL, 60637, USA                \\
    xqw3669@gmail.com}
\and
\IEEEauthorblockN{Xinye Sha}
\IEEEauthorblockA{
    \textit{Columbia University,        } \\
    New York, NY, 10027, USA               \\
    xinyesha1998@outlook.com}
\linebreakand 
\IEEEauthorblockN{Ying Chen}
\IEEEauthorblockA{
    \textit{Columbia University,} \\
    Sunnyvale, CA, 94086, USA \\
   yingchen.cheryl@gmail.com}
\and
\IEEEauthorblockN{Yuhui Jin}
\IEEEauthorblockA{
    \textit{Independent Researcher         } \\
    Pasadena, CA, 91106, USA               \\
    yuhuijin1995@gmail.com } 
\and
\IEEEauthorblockN{Qi Zheng}
\IEEEauthorblockA{
    \textit{Northeastern University} \\
    Boston, MA, 02115, USA\\    
    zheng.qi2@northeastern.edu}    
\and
\IEEEauthorblockN{Chang Yu$^*$}
\IEEEauthorblockA{
    \textit{Northeastern University} \\
    Boston, MA, 02115, USA\\
    chang.yu@northeastern.edu}
}

\begin{document}

\title{Application of an ANN and LSTM-based Ensemble Model for Stock Market Prediction
}

\maketitle

\begin{abstract}
Stock trading has always been a key economic indicator in modern society and a primary source of profit for financial giants such as investment banks, quantitative trading firms, and hedge funds. Discovering the underlying patterns within the seemingly volatile yet intrinsically structured economic activities has become a central focus of research for many companies. Our study leverages widely-used modern financial forecasting algorithms, including LSTM, ANN, CNN, and BiLSTM. We begin by comparing the predictive performance of these well-known algorithms on our stock market data, utilizing metrics such as R2, MAE, MSE, RMSE for detailed evaluation. Based on the performance of these models, we then aim to combine their strengths while mitigating their weaknesses, striving to construct a powerful hybrid model that overcomes the performance limitations of individual models.Through rigorous experimentation and exploration, we ultimately developed an LSTM+ANN model that breaks through prior performance bottlenecks, achieving promising and exciting results.

\end{abstract}


\section{Introduction}
In recent years, stock market trading has become a central element in the development of the global economy, drawing significant attention from investors, traders, and researchers alike. Accurately predicting the potential movement of stock prices is a critical aspect of stock analysis. Stocks represent a complex entity based on time series data and are influenced by many factors. For a long time, various linear and nonlinear traditional algorithms have been employed to decipher the nature of stocks and their potential trends. While these methods are highly advanced and have proven effective in many fields, they often need more stock analysis due to the inherent complexity and unpredictability of the stock market.

Addressing predictive challenges in time series domains has been a key focus within the machine learning community. Long Short-Term Memory (LSTM) models, initially proposed by Hochreiter and Schmidhuber in 1997\cite{202409.0662}, represent an auspicious approach for this task. Built upon recurrent neural network architectures, LSTM models effectively mitigate issues with vanishing and exploding gradients that frequently impair vanilla RNNs. This innovation has enabled significant advances across applications, including speech processing\cite{fan2024advanced}, handwriting recognition, and generative image modeling, where LSTMs have delivered state-of-the-art results. Researchers have found LSTM networks to yield significant predictive power within financial domains, leading to widespread adoption for time series forecasting in capital markets\cite{xu2022cardiac}.

Despite early theoretical limitations, artificial neural networks (ANNs) have exhibited comparable utility. Initially proposed in foundational work by Rosenblatt in 1958, ANNs encountered difficulties solving fundamental logical problems like XOR, forestalling their practical development. However, introducing backpropagation techniques in 1986, a significant milestone that refined approaches for training multilayer ANN models helped overcome these barriers, and catalyzed rapid progress. This discovery, which the field greatly appreciates, has led to ANNs demonstrating exceptional performance on tasks ranging from natural language processing to computer versions. Today, ANNs represent a flexible, widely used approach for financial forecasting\cite{tan2024enhanced}.

In our experiments, we employ LSTM and ANN models as base learners, harnessing two layers for the LSTM architecture. We then utilize linear regression as a meta-learner to ensemble their predictions. By synthesizing complementary strengths, we aim to transcend individual limitations and develop a robust, efficient hybrid model. Importantly, through rigorous comparative analysis, we ensure the reliability and validity of our findings, instilling confidence in the conclusions drawn. This approach is designed to advance the state-of-the-art in financial time series modeling and provide a solid foundation for future research.

In the upcoming sections, Section II will provide a brief overview of the dataset we are using, along with the knowledge background and architecture of the models employed. In Section III, we will compare our new hybrid model's performance against traditional models through a comprehensive parameter analysis that leaves no stone unturned, and experimental evaluation. Finally, in Section V, we will summarize and highlight the characteristics and value of this new ensemble model.

\section{Related Work}
This study builds upon the extensive body of research on LSTM and ANN models in the context of stock market prediction\cite{yang2024application}. Many of our methodologies, parameter comparisons, and research approaches are inspired by previous work in this field. Predicting stock prices, especially with the specific dataset we are using, presents significant challenges. Therefore, our aim is to explore a more practical approach by leveraging and refining the insights gained from prior research\cite{xu2024hybrid}.

Stock market prediction, especially using deep learning models like LSTM networks, has become vital because of its efficiency in financial prediction.\cite {zhu2021twitter}. Fischer and Krauss (2018) demonstrated the effectiveness of LSTM in financial market predictions, showing that these models can outperform traditional methods by learning from vast amounts of historical data and capturing long-term dependencies in stock prices\cite{fischer2018deep}.

ANN models, which are also called artificial networks, have been utilized widely in stock market prediction and non-linear prediction in the financial area.  After overcoming initial limitations, such as the inability to solve the XOR problem, ANNs gained prominence with the development of the backpropagation algorithm. This advancement enabled ANNs to effectively learn from historical stock data, leading to significant applications in financial forecasting. Researchers have shown that ANNs can capture complex patterns and trends within stock prices, making them a valuable tool in financial analysis and prediction\cite{zhang2001stock}.

Similarly, a pioneering work by Ni et al\cite{ni2024timeseries} demonstrates the superiority of transformer-based models in heart rate prediction, advancing both cardiovascular monitoring and broader time series forecasting applications. Inspired by the groundbreaking results of Ni et al\cite{ni2024timeseries}, our work extends the application of transformer-based models to a broader range of xxx and explores their potential in other areas such as financial forecasting, further demonstrating their versatility in capturing complex temporal dependencies across diverse domains.

Hu's research adeptly combines large-scale system analysis with AI-driven models, highlighting substantial improvements in traffic forecasting accuracy during chaotic events.\cite{hu2023artificial}
During our research, Hu's work offers crucial insights into the development of adaptive models for real-time traffic management\cite{hu2023artificial}.

SEO-LLM combines advanced techniques like Adaptive Data Augmentation and Synergistic Quantization, setting a new standard for optimizing large language models. This framework has great potential for social impact, enabling the use of high-performance language models in regions and industries with limited computational resources, which is highly beneficial for our research.\cite{202409.0662}. 

Ensemble models are a beneficial machine learning technology that combines the strengths of different models to improve predictive performance. Boosting and Stacking are two essential of ensemble models, with Stacking (Stacked Generalization) being a flexible approach where multiple independently trained base models' predictions are used as new features to let the metadata layer train and build new outputs based on base outputs, leveraging their diverse strengths and weaknesses to make more accurate predictions\cite{wolpert1992stacked}.

\section{Mythology}
In the following sections, we will provide a detailed introduction to our data sources, data processing methods, and the theoretical framework of our model. Initially, we will focus on explaining the data processing aspects Subsequently, we will gradually introduce the basic structure of our model, the overall design approach, and the underlying mathematical principles.

\subsection{Data Processing}
\subsubsection{Dataset Introduction}
We utilized a large dataset of trading data from S\&P 500 companies from New York Stock Market as our training data source. The dataset consists of 860 million trading records, including the companies' opening prices, closing prices, and daily high and low price indices. To facilitate analysis, the data has been split-adjusted, ensuring consistency across historical records and enabling more accurate model training.
\subsubsection{Data Cleaning}
Data cleaning is crucial before we start training our data to improve data quality and ensure accuracy, consistency, and completeness. We handled the missing values in the dataset by using median-based imputation to fill in the gaps\cite{bian2024relaxmore}. We also addressed and removed duplicate entries to ensure the dataset's integrity

\subsubsection{Standardization and Normalization}
Standardization transforms data, which will make mean 0 and standard deviation 1, effectively converting it to a standard normal distribution. This is particularly useful when the model assumes typically distributed data or features have different units or scales\cite{Shen2024Harnessing}. From the other side, finally, we will convert the value into a range[0, 1]. This method is proper when ensuring that all features contribute equally to distance measurements in algorithms like k-nearest neighbors or when dealing with data like image pixel values\cite{Weng202404}.

\subsubsection{Creating Time Series Data for Modeling}
This step employs a sliding window technique to convert the raw time series data into a format suitable for supervised learning. By defining a time step of 30, each input sequence (X) comprises 30 successive time points, with the corresponding output (y) being the value that immediately follows this sequence. This approach effectively captures the temporal dependencies within the data and improves the model's performance in learning from previous patterns for future prediction .

\subsubsection{Time Series Data Splitting}
To ensure the integrity of the time series forecasting process, TimeSeriesSplit is used to split data to train and test parts for further work. This method will train the value of historical data and evaluate it on future data, closely mirroring real-world forecasting scenarios. By selecting the last fold for testing, the performance of the model will be evaluated during training, providing an ideal evaluation of its predictive capabilities. We also have applied the method of Hu's research findings, significantly improving our traffic prediction models by integrating his advanced methodologies\cite{hu2023artificial}.

\subsection{Model}
\subsubsection{LSTM}
LSTM is a unique design model generated from the RNN model target to address training problems with high gradient \cite{li2023deception}. The core innovation of LSTM lies in its unique architecture, which combines the cell state and multiple gating mechanisms to manage and retain long-term dependencies effectively.

\paragraph{Cell State}
The cell state is the most critical component of LSTM. It functions like a conveyor belt, running through the entire network chain, and allows information to flow easily between different time steps. This capability enables the LSTM to retain important information over long sequences and update or discard it as necessary.

\paragraph{Gating Mechanisms}
LSTM regulates the flow of information through three key gates:

Forget Gate: This gate decides which information from the cell state should be discarded. It plays a crucial role in ensuring that irrelevant or outdated information does not persist as the model processes new inputs.

Input Gate: This gate determines what new information should be added to the cell state. First, it computes a control signal that regulates how much new information to incorporate. Then, it calculates a candidate value, which represents the potential information to be added to the cell state.

Output Gate: The output gate determines what information from the cell state should be used to generate the next hidden state. It outputs a value based on the current cell state and decides what should be passed on to the next time step, ensuring that the relevant information is propagated forward.

\paragraph{Cell State Update} 
The cell state is updated in each step, and the value is controlled by the retention of the previous cell state. In this way, it calculates layer by layer. 

The LSTM updates the cell state through linear summation, unlike the multiplicative approach used in traditional RNNs. This design choice is crucial for avoiding the vanishing gradient problem, which often prevents RNNs from effectively learning long-term dependencies. By updating the cell state additively, LSTMs help maintain stable gradients during backpropagation, enabling the network to learn and retain information over extended sequences more efficiently. The structure of LSTM is displayed as below in Fig. 1.

\begin{figure}
    \centering
    \includegraphics[width=0.8\linewidth]{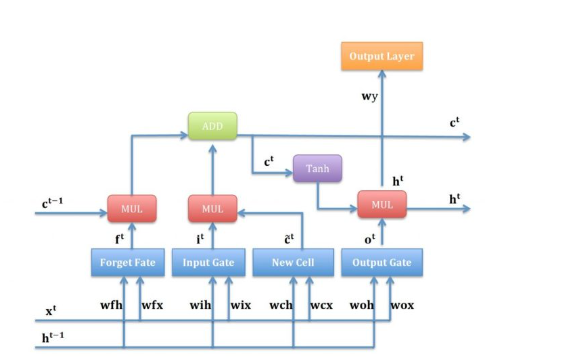}
    \caption{LSTM sturcture}
    \label{fig:enter-label}
\end{figure}

\subsubsection{ANN model}
The ann model is called an Artificial Neural Network. This model is created from the inspiration of biological neural structure, and it is designed to resolve complex data analysis cases\cite {Wu}. It consists of multiple simple processing units called neurons or nodes, interconnected through connections known as weights, forming a network structure\cite{gao2023autonomous}. ANN is built by the input layer and its hidden layer. The input layer will receive the data and send it to the next work with nodes in the hidden layer; the hidden layers are designed to be situated between their input and output layers; it will perform the majority of the computations and can have one or more layers, with the number of nodes adjusted based on the specific problem. Each node in its hidden layers applies a corresponding activation function to introduce nonlinearity, enabling the capture of more complex features. In the final part, the output layer will collect node info from different tasks and  generate a final decision, such as corresponding to the number of classes in a classification task or typically being a single node in a regression task.

\subsubsection{Ensemble Model Architecture and Mathematical Structure}

The ensemble model, a robust creation for predicting stock closing prices, combines the strengths of the LSTM and ANN models as we introduced. The predictions from these base models are then combined using a linear regression meta-model, ensuring more accurate and robust forecasts that you can rely on.

\paragraph{Base Models}

The ensemble is composed of two base models:

\begin{itemize}
    \item \textbf{Model LSTM:}The LSTM is a famous model designed for analyzing dependencies in the data of time series. The LSTM includes two years, every lawyer has 100 units, and it also has a following dropout layer to avoid overfitting. 

    \item \textbf{Model ANN:} ANN is a specially designed NN with two hidden parts. The first part of the layer is designed with 100 ReLU units activation, then in the next layer, there are 50 minutes to generate the final prediction.
\end{itemize}

\paragraph{Meta-Model}

The predictions from the LSTM and ANN models are combined using a stacking approach, with a linear regression model serving as the meta-model. The meta-model takes the predictions from the LSTM and ANN models, denoted as $\hat{y}_{\text{LSTM}}$ and $\hat{y}_{\text{ANN}}$, as input features.

\begin{equation}
\hat{y}_{\text{meta}} = \beta_0 + \beta_1 \hat{y}_{\text{LSTM}} + \beta_2 \hat{y}_{\text{ANN}} + \epsilon
\end{equation}

Here, $\hat{y}_{\text{meta}}$ is the final ensemble prediction, $\beta_0$ is the intercept, $\beta_1$ and $\beta_2$ are the coefficients learned by the linear regression model, and $\epsilon$ represents the error term.

\paragraph{Data Processing and Model Evaluation}

The time series data is first normalized using a MinMaxScaler, and its range is from 0 to 1. Training and testing datasets are all created from it. The LSTM and ANN models are built independently from their training dataset\cite{article}. The meta-layer will then determine the financial result, which the linear regression model builds .

Designed parameters will evaluate the result of the ensemble, and finally, all these results will be compared with the origin value. 

\section{Evaluation}
\subsection{Evaluation Metric}
We have implemented a few variables for the model performance evaluation. The $R^2$ is designed to evaluate how well the model fits the origin data. The MAE (Mean et al.) is the variable that calculates the average absolute diff between actual data and predicted data\cite{kral2013crack}, where smaller values indicate lower error\cite{Ma_2024}. MSE targets the average square difference between the predicted and actual values. Finally, the RMSE  will calculate the square root of the value generated from the MSE, reflecting the scale of the prediction error, where smaller values suggest higher predictive accuracy.
\subsection{Experiment}
Stock prediction is a tough topic because of its complexity and no-linear structure. News events and market noise can influence short-term movements in the broader market. Additionally, stock trading data contains significant noise and uncertainty stemming from the market's inherent unpredictability. In the time series context, stock prices exhibit non-stationary characteristics, meaning their statistical properties, such as mean and variance, can change over time. Moreover, a wide range of technical indicators, fundamental data, and macroeconomic factors can substantially impact stock prices. To comprehensively assess model performance, we employ several evaluation metrics, including four variables we have mentioned in the previous part\cite{white2022measuring}.

Our study compared the performance of various high-performing classical models, including CNN, ANN, LSTM, BiLSTM, RNN, LSTM+RNN, and ANN+CNN. Applied method introduced by Ni et. \cite{ni2024timeseries}, we also implemented a range of models from traditional ARIMA to advanced transformer architectures, focusing on the high-performing PatchTST model.
Our novel hybrid model demonstrated solid overall performance through this comprehensive comparison.
\begin{table}[h!]
\centering
\caption{Performance of Various Models}
\begin{tabular}{@{}p{3.5cm}@{\hspace{3mm}}c@{\hspace{3mm}}c@{\hspace{3mm}}c@{\hspace{3mm}}c@{}}
\hline
\textbf{Model} & \textbf{R\textsuperscript{2}} & \textbf{MAE} & \textbf{MSE} & \textbf{RMSE} \\
\hline
BiLSTM        & -0.0784  & 46.3969 & 11345.6214 & 106.5158 \\
CNN           &  0.1918  & 51.1716 &  8502.5717 &  92.2094 \\
ANN           &  0.4098  & 46.9158 &  6209.0942 &  78.7978 \\
LSTM          &  0.2717  & 42.3914 &  7662.1046 &  87.5334 \\
RNN           & -0.0895  & 46.8068 & 11461.7849 & 107.0597 \\
ARIMA         & -0.0062  & 47.6285 & 10584.8683 & 102.8828 \\
\hline
BLSTM + CNN   &  0.1468  & 45.8925 &  8975.8217 &  94.7408 \\
LSTM + CNN    & -0.1109  & 52.7154 & 11687.5362 & 108.1089 \\
\textbf{ANN + CNN (Stacking)} & \textbf{0.4631} & \textbf{42.3928} & \textbf{5648.1410} & \textbf{75.1541} \\
\textbf{LSTM + ANN (Final)} & \textbf{0.5375} & \textbf{37.7829} & \textbf{4865.2330} & \textbf{69.7512} \\
\hline
\end{tabular}
\label{tab:model_performance}
\end{table}
A detailed comparison of model performance shows that all the models exhibited significant limitations when applied to stock analysis. ANN, LSTM, and CNN performed slightly better than other deep-learning models\cite{chen2024taskclip}. However, due to their generally poor performance, none of these models, except for ANN, demonstrated sufficient practical value. After applying ensemble techniques, we observed substantial performance improvements in our constructed ANN-CNN and LSTM+ANN models, with the failure performance of the BiLSTM+CNN and LSTM+CNN models.

The performance of various models, as summarized in Table \ref{tab:model_performance}, indicates the superior predictive capability of the LSTM+ANN model compared to other models evaluated in this study. The LSTM+ANN model achieved the highest $R^2$ value of 0.5375, which indicates that it explains the variance in the data more effectively than the other models. This is a significant improvement over other models, such as the ANN model ($R^2 = 0.4098$) and the CNN model ($R^2 = 0.1918$), clearly demonstrating the good result of the LSTM+ANN ensemble model in our training.

In terms of error metrics, the LSTM+ANN model also outperforms its counterparts. It achieved the lowest MAE (Mean et al.) at 37.7829, indicating that, on average, its predictions are closer to the actual values than those of other models. For example, the ANN and LSTM models have MAE values of 46.9158 and 46.4096, respectively, which are significantly higher than those of LSTM+ANN.

Moreover, the LSTM+ANN model's MSE (Mean Squared Error) is the lowest among all models, at 4865.2330, and its RMSE (Root Mean Squared Error) is also the lowest, at 69.7512. These metrics show that the LSTM+ANN model minimizes the average prediction error and effectively reduces the impact of more significant errors, which is critical in stock prediction tasks where large deviations can lead to significant financial consequences\cite{bian2022optimal}.

Based on the data from the table, it is evident that among all the deep learning models we tested, the ANN model performed the best. However, our LSTM+ANN ensemble model, which is built on the ANN model, demonstrated over a 20\% improvement in performance, successfully breaking through the performance bottlenecks of all the original models and delivering a remarkable enhancement in results.

Overall, the combination of LSTM's ability to capture temporal dependencies and ANN's capacity to model complex nonlinear relationships has resulted in a model that excels across all key performance metrics, making the LSTM+ANN model the most robust and reliable model in this study for stock price prediction.

\section{Conclusion}

Compared to other financial tasks like Fraud Protection and Credit Prediction, Stock Prediction poses a far more significant challenge. This issue arises from the sheer volume of time-series features and the influence of breaking news and unforeseen events on the stock market, making accurate predictions exceedingly tricky.

Our experiments with native Deep Learning models\cite{zhou2024optimizing}, even the most advanced ones tailored for stock market analysis, have revealed significant performance issues. The error rates are consistently high, and R-squared values often fall below 0.5, indicating underwhelming performance. Models like LSTM and ANN have yet to meet our expectations\cite{ji2023prediction}.

However, by optimizing and integrating our models through an Ensemble Model Stacking architecture, we crafted an innovative hybrid model that merges LSTM and ANN. This new model marks a revolutionary performance breakthrough, achieving an R-squared value exceeding 0.5 for the first time. It also surpasses the current leading model, ANN, by over 20\%. A detailed comparison and discussion of specific parameters can be found in the Evaluation section.

The remarkable outcomes of this experiment not only represent a significant performance enhancement but point to a promising avenue for further research. We have overcome a technical bottleneck by utilizing Ensemble Model Stacking and can now aim for more extreme, powerful, and efficient stock prediction models\cite{luo2023towards}. We will continue to explore this path, anticipating that our findings will substantially impact social and economic development.

\bibliographystyle{plain}
\bibliography{ref}

\begin{thebibliography}{10}

\bibitem{bian2022optimal}
Wanyu Bian, Yunmei Chen, and Xiaojing Ye.
\newblock An optimal control framework for joint-channel parallel mri reconstruction without coil sensitivities.
\newblock {\em Magnetic Resonance Imaging}, 2022.

\bibitem{bian2024relaxmore}
Wanyu Bian, Albert Jang, and Fang Liu.
\newblock Improving quantitative mri using self-supervised deep learning with model reinforcement: Demonstration for rapid t1 mapping.
\newblock {\em Magnetic Resonance in Medicine}, 2024.

\bibitem{chen2024taskclip}
Hanning Chen, Wenjun Huang, Yang Ni, Sanggeon Yun, Fei Wen, Hugo Latapie, and Mohsen Imani.
\newblock Taskclip: Extend large vision-language model for task oriented object detection.
\newblock {\em arXiv preprint arXiv:2403.08108}, 2024.

\bibitem{article}
Bo~Dang, Danqing Ma, Shaojie Li, Zongqing Qi, and Elly Zhu.
\newblock Deep learning-based snore sound analysis for the detection of night-time breathing disorders.
\newblock {\em Applied and Computational Engineering}, 76:109--114, 07 2024.

\bibitem{fan2024advanced}
Xiaojing Fan, Chunliang Tao, and Jianyu Zhao.
\newblock Advanced stock price prediction with xlstm-based models: Improving long-term forecasting.
\newblock {\em Preprints}, (2024082109), August 2024.

\bibitem{fischer2018deep}
Thomas Fischer and Christopher Krauss.
\newblock Deep learning with long short-term memory networks for financial market predictions.
\newblock {\em European Journal of Operational Research}, 270(2):654--669, 2018.

\bibitem{gao2023autonomous}
Longsen Gao, Giovanni Cordova, Claus Danielson, and Rafael Fierro.
\newblock Autonomous multi-robot servicing for spacecraft operation extension.
\newblock In {\em 2023 IEEE/RSJ International Conference on Intelligent Robots and Systems (IROS)}, pages 10729--10735. IEEE, 2023.

\bibitem{hu2023artificial}
Tiechuan Hu, Wenbo Zhu, and Yuqi Yan.
\newblock Artificial intelligence aspect of transportation analysis using large scale systems.
\newblock In {\em Proceedings of the 2023 6th Artificial Intelligence and Cloud Computing Conference}, pages 54--59, 2023.

\bibitem{ji2023prediction}
Yuelyu Ji, Yuhe Gao, Runxue Bao, Qi~Li, Disheng Liu, Yiming Sun, and Ye~Ye.
\newblock Prediction of covid-19 patients’ emergency room revisit using multi-source transfer learning.
\newblock In {\em 2023 IEEE 11th International Conference on Healthcare Informatics (ICHI)}, pages 138--144. IEEE, 2023.

\bibitem{kral2013crack}
Zachary Kral, Walter Horn, and James Steck.
\newblock Crack propagation analysis using acoustic emission sensors for structural health monitoring systems.
\newblock {\em The Scientific World Journal}, 2013(1):823603, 2013.

\bibitem{li2023deception}
Panfeng Li, Mohamed Abouelenien, Rada Mihalcea, Zhicheng Ding, Qikai Yang, and Yiming Zhou.
\newblock Deception detection from linguistic and physiological data streams using bimodal convolutional neural networks.
\newblock {\em arXiv preprint arXiv:2311.10944}, 2023.

\bibitem{202409.0662}
Xinjin Li, Yu~Ma, Yangchen Huang, Xingqi Wang, Yuzhen Lin, and Chenxi Zhang.
\newblock Integrated optimization of large language models: Synergizing data utilization and compression techniques.
\newblock {\em Preprints.10.20944/preprints202409.0662.v1}, September 2024.

\bibitem{luo2023towards}
Zhimeng Luo, Yuelyu Ji, Abhibha Gupta, Zhuochun Li, Adam Frisch, and Daqing He.
\newblock Towards accurate and clinically meaningful summarization of electronic health record notes: A guided approach.
\newblock In {\em 2023 IEEE EMBS International Conference on Biomedical and Health Informatics (BHI)}, pages 1--5. IEEE, 2023.

\bibitem{Ma_2024}
Danqing Ma, Shaojie Li, Bo~Dang, Hengyi Zang, and Xinqi Dong.
\newblock Fostc3net: A lightweight yolov5 based on the network structure optimization.
\newblock {\em Journal of Physics: Conference Series}, 2824(1):012004, August 2024.

\bibitem{ni2024timeseries}
Haowei Ni, Shuchen Meng, Xieming Geng, Panfeng Li, Zhuoying Li, Xupeng Chen, Xiaotong Wang, and Shiyao Zhang.
\newblock Time series modeling for heart rate prediction: From arima to transformers.
\newblock {\em arXiv preprint arXiv:2406.12199}, 2024.

\bibitem{Shen2024Harnessing}
Xinyu Shen, Qimin Zhang, Huili Zheng, and Weiwei Qi.
\newblock {Harnessing XGBoost for robust biomarker selection of obsessive-compulsive disorder (OCD) from adolescent brain cognitive development (ABCD) data}.
\newblock In Pier~Paolo Piccaluga, Ahmed El-Hashash, and Xiangqian Guo, editors, {\em Fourth International Conference on Biomedicine and Bioinformatics Engineering (ICBBE 2024)}, volume 13252, page 132520U. International Society for Optics and Photonics, SPIE, 2024.

\bibitem{tan2024enhanced}
Lianghao Tan, Shubing Liu, Jing Gao, Xiaoyi Liu, Linyue Chu, and Huangqi Jiang.
\newblock Enhanced self-checkout system for retail based on improved yolov10.
\newblock {\em arXiv preprint arXiv:2407.21308}, 2024.

\bibitem{Weng202404}
Yijie Weng and Jianhao Wu.
\newblock Big data and machine learning in defence.
\newblock {\em International Journal of Computer Science and Information Technology}, 16(2), 2024.

\bibitem{white2022measuring}
Veronica~M White and Joel~M Hunt.
\newblock Measuring how relatively ‘good’a hot-spot map is: A summary of current metrics.
\newblock In {\em Proceedings of the IISE Annual Conference \& Expo 2022}, 2022.

\bibitem{wolpert1992stacked}
David~H Wolpert.
\newblock Stacked generalization.
\newblock {\em Neural networks}, 5(2):241--259, 1992.

\bibitem{Wu}
Di~Wu.
\newblock The effects of data preprocessing on probability of default model fairness.
\newblock {\em World Journal of Advanced Engineering Technology and Sciences}, Aug 2024.

\bibitem{xu2024hybrid}
Yongshun Xu, Shuo Han, Dayang Wang, Ge~Wang, Jonathan~S Maltz, and Hengyong Yu.
\newblock Hybrid u-net and swin-transformer network for limited-angle cardiac computed tomography.
\newblock {\em Physics in Medicine \& Biology}, 69(10):105012, 2024.

\bibitem{xu2022cardiac}
Yongshun Xu, Asif Sushmit, Qing Lyu, Ying Li, Ximiao Cao, Jonathan~S Maltz, Ge~Wang, and Hengyong Yu.
\newblock Cardiac ct motion artifact grading via semi-automatic labeling and vessel tracking using synthetic image-augmented training data.
\newblock {\em Journal of X-Ray Science and Technology}, 30(3):433--445, 2022.

\bibitem{yang2024application}
Yutian Yang, Hongjie Qiu, Yulu Gong, Xiaoyi Liu, Yang Lin, and Muqing Li.
\newblock Application of computer deep learning model in diagnosis of pulmonary nodules.
\newblock {\em arXiv preprint arXiv:2406.13205}, 2024.

\bibitem{zhang2001stock}
G~Peter Zhang.
\newblock Stock market prediction of s\&p 500 index using artificial neural networks.
\newblock {\em Neural Networks}, 12(4-5):865--877, 2001.

\bibitem{zhou2024optimizing}
Chang Zhou, Yang Zhao, Jin Cao, Yi~Shen, Jing Gao, Xiaoling Cui, Chiyu Cheng, and Hao Liu.
\newblock Optimizing search advertising strategies: Integrating reinforcement learning with generalized second-price auctions for enhanced ad ranking and bidding.
\newblock {\em arXiv preprint arXiv:2405.13381}, 2024.

\bibitem{zhu2021twitter}
Wenbo Zhu and Tiechuan Hu.
\newblock Twitter sentiment analysis of covid vaccines.
\newblock In {\em 2021 5th International Conference on Artificial Intelligence and Virtual Reality (AIVR)}, pages 118--122, 2021.

\end{thebibliography}

\end{document}